\documentclass[aps,prb, twocolumn,superscriptaddress,amsmath,showpacs,showkeys]{revtex4}

\usepackage[latin1]{inputenc}
\usepackage{graphicx}
\usepackage{dcolumn}
\usepackage{bm}
\usepackage{setspace}

\newcommand {\etal} {\textit{et al.}}

\newcommand {\Tz} {\ensuremath{T_{0}}}

\newcommand {\CeRuAl} {CeRu$_{2}$Al$_{10}$}

\begin{document}

\title{Long-range order and low-energy magnetic excitations in \CeRuAl.}

\author{Julien Robert}
\email[e-mail address: ]{julien.robert@cea.fr}
\author{Jean-Michel Mignot}
\author{Gilles Andr\'{e}}
\affiliation{Laboratoire L\'{e}on Brillouin, CEA-CNRS, CEA/Saclay, 91191 Gif sur Yvette (France)}

\author{Takashi Nishioka}
\author{Riki Kobayashi}
\author{Masahiro Matsumura}
\affiliation{Graduate School of Integrated Arts and Science, Kochi University, Kochi 780-8520 (Japan)}

\author{Hiroshi Tanida}
\author{Daiki Tanaka}
\author{Masafumi Sera}
\affiliation{Department of Quantum Matter, ADSM, Hiroshima University, Higashi-Hiroshima, 739-8530 (Japan)}

\date{\today}

\begin{abstract}
The nature of the unconventional ordered phase occurring in \CeRuAl\ below $\Tz = 27$ K was investigated by neutron scattering. Powder diffraction patterns show clear superstructure peaks corresponding to forbidden $(h+k)$-odd reflections of the $Cmcm$ space group. Inelastic neutron scattering experiments further reveal a pronounced magnetic excitation developing in the ordered phase at an energy of 8 meV. 
\end{abstract}



\maketitle

The low-temperature behavior of intermetallic cerium compounds can be broadly typified in terms of the competition between several interaction channels (intra-atomic couplings, on-site Coulomb repulsion, hybridization between local $f$-electron states and itinerant conduction-band states), forming the basis of the well-known and highly successful Anderson model.\cite{Anderson'61} 
However, there has also been continued interest in Ce-based materials which do not seem to fit into this general framework. Among those are, for instance, the ``Kondo insulators'', as well as various compounds exhibiting multipole ordering \cite{Kuramoto'09} or other types of elusive ``hidden order'' transitions. One example of such unconventional ordering properties has been discovered very recently by Strydom \cite{Strydom'09} in the ternary  compound \CeRuAl.

\CeRuAl\ is an YbFe$_2$Al$_{10}$-type orthorhombic compound belonging to the $Cmcm$ space group, with room-temperature lattice constants $a =$ 9.1272 \AA, $b =$ 10.282 \AA, and $c =$ 9.1902 \AA. It has been described as a ``cage'' crystal structure, in which Ce atoms are separated from each other by an exceptionally large distance of 5.2 \AA. From the lattice constants, the Ce valence state was estimated to be close to 3+. The transport properties below room temperature are indicative of a gap in the electronic structure,\cite{Strydom'09} although the Hall effect still suggests a dominant metallic character. In this regime, the material exhibits considerable magnetic anisotropy ($a$: easy axis, $b$: hard axis).\cite{Nishioka'09,Tanida'10.1,Tanida'10.2a} Upon application of pressure, the system rapidly changes, first to a Kondo insulator, then to a metal above 5 GPa.\cite{Nishioka'09} The striking feature of this compound is the phase transition taking place at  $\Tz = 27$ K, which causes pronounced anomalies in various physical properties. Whereas the origin of this transition remains highly controversial, there is growing evidence that it cannot reduce to a conventional ordering of local Ce magnetic moments. 
The  transition temperature is far too high in view of the large Ce--Ce distance and, more specifically, of the magnetic ordering temperature of 16.5 K found in GdRu$_2$Al$_{10}$.\cite{Nishioka'09} The drop in the magnetic susceptibility below \Tz\ occurring for \textit{all three} magnetic field orientations $H \parallel a, b, c$, with an exponential behavior $\chi = \chi_0 + A \exp(-\Delta/T)$ and $\Delta \sim 100$ K, is also difficult to reconcile with the behavior expected for an antiferromagnet.\cite{Nishioka'09,Tanida'10.0} Finally,  $^{27}$Al NQR/NMR experiments did not find the splitting of peaks below \Tz\ expected for a static order of Ce magnetic moments.\cite{Matsumura'09} 
 
Alternative mechanisms such as charge- or spin-density-wave formation also have serious shortcomings.\cite{Nishioka'09,Matsumura'09} Recently, Tanida et al.\cite{Tanida'10.0,Tanida'10.1} suggested the formation of a Ce dimer within the $a,c$ plane of the crystal, bearing some similarities to a spin-Peierls transition, but conclusive experimental evidence is still missing. To assess these competing interpretations, one first needs to characterize the order forming below \Tz\  and to determine the nature of the low-energy excitations. In this work, we have addressed the latter questions by performing the first neutron powder diffraction (NPD) and inelastic neutron scattering (INS) experiments on this compound. The results clearly show the appearance of weak superstructure peaks reflecting a long range order (LRO) setting in below the transition at  \Tz, as well as a pronounced magnetic excitation developing in the ordered phase at an energy of 8 meV.

Neutron scattering experiments were carried out  at Orph\'ee-LLB in Saclay. The main part of the sample, consisting of  23.7 g of annealed  \CeRuAl\ powder, was characterized by x-ray diffraction and found to be single-phase, apart from one very weak reflection from an unknown foreign constituent. An extra 14 g of compound was prepared by crushing tiny single-crystal pieces grown by the Al self-flux method. The latter material was of higher quality and contained no extra phase. It was  used alone for the NPD measurements on the two-axis diffractometer G4-1 (800-cell position-sensitive detector). The sample was contained inside a thin-walled vanadium cylinder, 10 mm in diameter, and studied at an incident wavelength of 2.423 \AA. A pyrolytic graphite (PG) filter was placed in the incident beam to suppress higher-order contaminations.  The data analysis was performed using the Rietveld refinement program \textsc{FullProf}.\cite{fullprof'93,fullprof'01} All available sample material was then combined for the INS experiments, performed on the triple-axis spectrometers 2T (thermal beam) and 4F2 (cold source), using PG 002 monochromators and analyzers. Spectra were recorded in the constant-$k_f$ mode at final neutron energies $E_f$ of 14.7 and 5.0 meV, with PG and Be filters, respectively, inserted in the scattered beam.

 \begin{figure}  
	\includegraphics [width=0.95\columnwidth, angle=0] {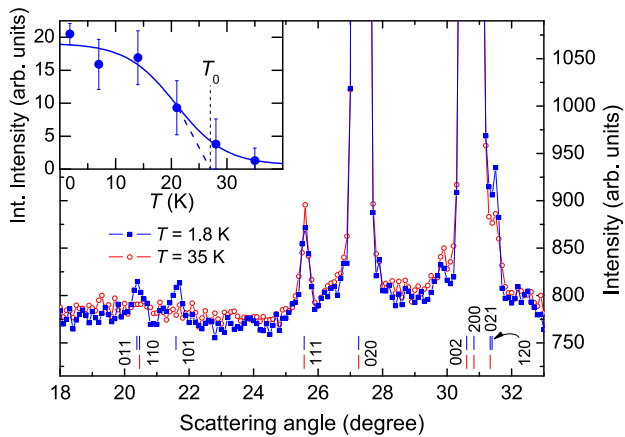}
	\caption{\label{diff_patt} (Color online) Low-angle region of the neutron diffraction patterns of \CeRuAl\ at $T_\mathrm{min} = 1.8$ K and $T = 35$ K $> \Tz$. The two series of ticks below the trace denote the reflections observed at 1.8 K (upper set) and 35 K (lower set).	(Inset): temperature dependence of the integrated intensity of the 101 peak.}
\end{figure}

NPD patterns were measured at 6 different temperatures, $T = 1.8, 7, 14, 21, 28$, and 35 K, both below and above the ordering temperature \Tz. The nuclear Bragg reflections in the paramagnetic phase could be indexed in the crystal structure reported by Tursina \etal\ \cite{Tursina'05} with a Bragg reliability factor $R_{B} = 9.8$.\footnote{The quality of this refinement was somewhat reduced by the presence of crystallites in the measured powder.} No sizable change in the lattice constants was detected at the transition, in agreement with previous thermal expansion data.\cite{Tanida'10.0} Below \Tz, weak additional reflections appear in the low-angle range, as is clearly seen in Fig.~\ref{diff_patt}. They correspond to the scattering angles calculated for the 011, 101 and 120 reflections, which are forbidden in the high-temperature $Cmcm$ space group. This indicates that the selection rule $h+k = 2n$ due to the centering of the $(a,b)$ orthorhombic faces is relaxed in the ordered state.  Within experimental accuracy, the superstructure peaks exhibit no extra broadening in comparison with the neighboring nuclear peaks, which indicates that the order is long-range. The temperature dependence plotted in the inset in Fig.~\ref{diff_patt} shows that the intensity of the 101 peak vanishes around the transition temperature $\Tz = 27$ K. 

Because of the small number, low intensity, and limited angular range of the observed superstructure reflections, it cannot be decided from these NPD data alone whether they are magnetic or nuclear in origin. In the former case, the data can be ascribed to a collinear antiferromagnetic (AFM) structure with wavevector $\bm{k}=(1,0,0)$. The refinement yields a Ce magnetic moment comprised between 0.3 and 0.5 $\mu_B$/Ce, depending on its assumed orientation. The main problem with this interpretation is to reconcile the existence of this ordered magnetic component with the splitting of the NQR lines observed by Matsumura et al.\cite{Matsumura'09} 
Alternatively, the peaks could arise from a symmetry-breaking structural transition. In Ref.~[\onlinecite{Matsumura'09}], it was argued that, since the transition appears to be second order, the symmetry of the ordered state should correspond to one of the two subgroups $Amm$2 or $Pmmn$ of the $Cmcm$ space group. The observation of the 101 peak in the present measurements rules out the former subgroup (equivalent to $Cm2m$ for the original labeling of axes), but $Pmmn$ is consistent with all observed superstructure peaks. In this case, the existence of the 101 peak would require a displacement of Al atoms, since its intensity is insensitive to the position parameter ($b$ component) of the Ce or Ru sites.\cite{Cousson'priv} A similar picture was proposed in a recent theoretical work by Hanzawa,\cite{Hanzawa'10.1} based on a spin-Peierls (dimerization) transition along 1D zig-zag chains, which may be accompanied by a 3D order of the bonds connecting Ce--Ce dimers.\cite{Hanzawa'10.2a} On the other hand, such a model cannot account for the weak static field ($\approx 30$ G) observed below \Tz\ in recent $\mu ^+$SR measurements.\cite{Kambe'10} The estimated value of the ordered moment is quite small, about 10$^{-2}\mu_{B}$/Ce (undetectable in the NPD experiments), and it is thus presently unclear whether the signals observed by neutrons and muons have a common origin.

 \begin{figure}  
	\includegraphics [width=1.0\columnwidth, angle=0] {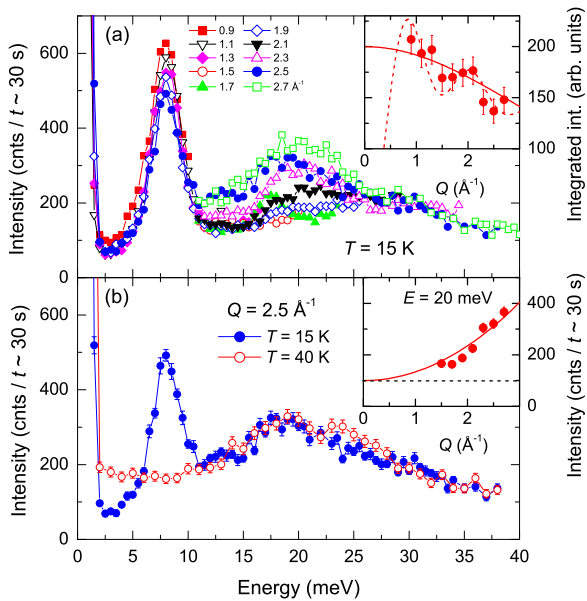}
	\caption{\label{q_dep} (Color online) (a) INS spectra of \CeRuAl\ at $T = 15$ K measured at fixed final energy $E_f =$ 14.7 meV for different momentum transfers (for clarity, some data sets have been restricted to $E > 10$ meV); inset: $Q$ dependence of the integrated intensity of the low-energy peak. (b) Comparison of spectra below (15 K ) and above (40 K) \Tz; inset: $Q$ dependence of the intensity at $E = 20$ meV; solid and dashed lines denote calculations for a single-ion Ce form factor or a Ce-dimer structure factor, respectively.}
\end{figure}

INS spectra at the minimum temperature of 14 K were measured at a fixed final energy $E_f =$ 14.7 meV. The data for different momentum transfers $Q$ between 0.9 and 2.7 \AA$^{-1}$ are collected in Fig.~\ref{q_dep}(a). At all $Q$ values, a pronounced excitation exists close to 8 meV, along with broader features at higher energies, in particular a maximum occurring around 20 meV for the larger $Q$ values. The integrated intensity of the lower peak decreases with increasing $Q$, approximately following the Ce$^{3+}$ magnetic dipole form factor, which shows that the origin of this excitation is primarily magnetic. As represented in the inset of Fig.~\ref{q_dep}(a), comparable, or slightly better agreement is obtained if one takes into account the structure factor for a Ce dimer with a Ce--Ce spacing of 5.2 \AA. In contrast, the upper peak is enhanced at large momentum transfers: the $Q$ dependence of the peak intensity at $E = 20$ meV, plotted in the inset of Fig.~\ref{q_dep}(b), is roughly consistent with a $Q^2$ law, indicative of nuclear phonon scattering. These different origins are further substantiated by the temperature evolution of the inelastic signal displayed in Fig.~\ref{q_dep}(b) for $Q = 2.5$ \AA$^{-1}$. The two curves plotted for $T =$ 15 and 40 K show practically no difference at energies above 12 meV, whereas the 8-meV peak occurring at 15 K is completely suppressed in the disordered state.
 
 \begin{figure}  
	\includegraphics [width=0.80\columnwidth, angle=0] {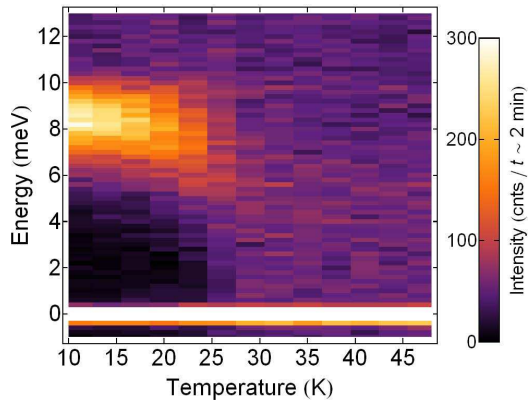}
	\caption{\label{int_map} (Color online) Intensity map of the INS spectra of \CeRuAl\ measured at $E_f =$ 5.0 meV and for a momentum transfer of 1.5 \AA$^{-1}$, showing the spectral weight associated to the the 8-meV peak in the ordered state, and its transfer to a wide energy window above $\Tz = 27$ K. }
\end{figure}

The energy region below 13 meV was studied with a higher resolution using cold neutrons of final energy $E_f =$ 5.0 meV. The results are summarized in the intensity map of Fig.~\ref{int_map} for temperatures between 10 and 47 K. At low temperature, one notes the strong intensity concentrated at the position of the peak just above 8 meV, whereas there exists  essentially no  signal below 5 meV. This clearly indicates the existence of a gap in the spectrum of magnetic excitations. With increasing temperature, the excitation broadens and shifts to lower energies, while a featureless background, extending all the way to zero energy, appears above \Tz. 

 \begin{figure}  [t] 
	\includegraphics [width=0.90\columnwidth, angle=-0] {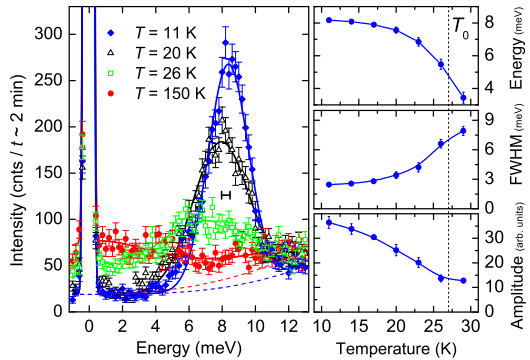}
	\caption{\label{t_dep} (Color online)  Left frame: selected constant-$Q$ scans ($Q =  1.5$ \AA$^{-1}$) measured in \CeRuAl\ at temperatures between 11 K and 150 K, for a fixed final neutron energy of 5.0 meV; solid lines are fits using quasielastic Lorentzian and inelastic Gaussian lineshapes (see text); dashed lines represent the estimated background for the two extreme temperatures; the ``H'' mark denotes experimental resolution at 8 meV. Right frames (top to bottom): energy, full width at half maximum, and amplitude of the inelastic peak.}
\end{figure}

Figure \ref{t_dep} summarizes the analysis of the low-energy data. In the main frame, constant-$Q$ scans are plotted for different temperatures below and above \Tz. With increasing temperature, the intensity of the inelastic peak decreases steadily while its linewidth increases, and its position shifts to lower energies. As temperature approaches \Tz, significant intensity appears at lower energy, resulting in the suppression of the spin gap. This signal extends over practically the whole measured energy range and, above \Tz, produces a nearly temperature-independent intensity up to at least $T = 150$ K. The transfer of spectral weight, upon heating, from the 8-meV excitation to a very broad signal roughly centered at zero energy can be represented by two magnetic spectral components, one inelastic and one quasielastic, with opposite temperature dependences. The experimental spectra were accordingly fitted to a scattering function $S_{\text{mag}}$ comprised, besides the incoherent elastic peak (resolution-limited Gaussian) and a sloping background (tail of the phonon signal seen on Fig.~\ref{q_dep}) weighted by the detailed balance factor, of two magnetic spectral functions:
one broad quasielastic Lorentzian and one Gaussian peak centered at finite energy. The former spectral shape was taken phenomenologically, since the experimental data indicate a rather complex profile, with a total width exceeding by a factor of  5 the instrumental resolution. The parameters from the fit are plotted as a function of temperature in the three right-hand side frames in the figure. The spectral weight derived from the integrated intensity of the peak does not change significantly up to 20 K. 

During the reviewing of this paper, Khalyavin et al.\cite{Khalyavin'10a} reported time-of-flight NPD measurements on the same compound. The data are closely similar to those presented above. However, the authors claim that their refinements unambiguously establish the magnetic origin of the superstructure peaks, associated with an AFM wave vector $\bm{k}=(1,0,0)$. The reliability of this conclusion is difficult to assess from the data shown in the report but, if it were to be confirmed, it would directly contradict the interpretation of the previous NQR measurements,\cite{Matsumura'09} taken to rule out long-range magnetic order, and thus reinforce the unconventional character of the ordered state in \CeRuAl. 
From the present INS study, this ordering was seen to be accompanied by the appearance of a well-defined, predominantly magnetic excitation at  8 meV. A crystal-field (CEF) transition can be ruled out because of the drastic suppression of the neutron peak occurring above \Tz. Indeed, the measured entropy indicates only one doublet to exist at least up to 100 K ($\approx$ 9 meV). Furthermore, the width of the inelastic peak is quite large, so that a similarly broadened, and thus readily observable, quasielastic signal would have been expected from fluctuations within a Kramers doublet CEF ground state. A central role played by Ce-Ce exchange interactions was previously inferred from the fact that the transition temperature is strongly suppressed upon substitution of Ce by La.\cite{Tanida'10.0,Kondo'10} Specific-heat measurements \cite{Nishioka'09} also showed that an entropy of $0.7R\ln{2}$ is recovered at \Tz ($R\ln{2}$ at about 100 K), which implies that the degeneracy of the lowest Ce$^{3+}$ CEF doublet is lifted below the transition. The observed behavior could be qualitatively explained in a singlet-ground-state picture. Magnetic excitations at finite energies, arising from a (many-body) singlet ground  state, and changing to a broad quasielastic response with increasing temperature, were indeed reported for some mixed-valence (CeNi, YbAl$_3$) or Kondo-insulator (SmB$_6$, YbB$_{12}$) compounds. In such cases, however, the change in the magnetic response as a function of temperature occurs through a gradual crossover rather than a phase transition. Moreover, there is no evidence here, from the lattice constants, for a significant valence change occurring at \Tz. In the recent work of Refs.~[\onlinecite{Tanida'10.0,Tanida'10.1,Tanida'10.2a}], it was suggested that a spin-pairing transition takes place at \Tz, producing a singlet ground state and a spin gap of the order of 100 K. In such a case, the excitation could correspond to a singlet-triplet (S-T) transition from the dimer ground state. Its broadening, observed even in the measurements at $E_f =$ 5.0 meV, may be intrinsic (magnetic fluctuations) or due to a dispersion of the mode, powder-averaged in $Q$ space. However, the large exchange interaction ($J \sim 8$ meV) required by a S-T model seems difficult to reconcile with the first-neighbor Ce distance of 5.2 \AA\ in this material. If, on the other hand, long-range AFM order does occur below \Tz, with Ce moments aligned along the $c$ axis as claimed in Ref.~[\onlinecite{Khalyavin'10a}], the excitation could be ascribed to a magnon branch with an anisotropy gap, but the origin of the large gap value, and comparatively weak overall dispersion, is not obvious, especially since the strong magnetic anisotropy found in the paramagnetic state was shown to have its easy axis oriented along the $a$ direction.\cite{Tanida'10.2a}

In conclusion, this neutron scattering study of \CeRuAl\ revealed a  well-defined superstructure corresponding to the order formed below \Tz. This superstructure may results either from small periodic displacements of Al (and possibly other) atoms---in which case the results place further constraints on the symmetry-breaking suggested from previous NQR measurements, or possibly from an AFM order of weak Ce magnetic moments. The spectrum of low-energy excitations in this LRO state, characterized by a spin-gap and a predominantly magnetic peak at $E \approx 8$ meV, may be ascribed to singlet-triplet excitations from dimerized pairs of Ce ions, but an AFM magnon cannot be ruled out in case AFM order does exist. Additional neutron experiments on a single crystal and/or using polarized neutrons are needed to answer these questions.

We thank F. Maignen for technical support and A. Cousson and P. A. Alekseev for helpful comments.

\end{document}